\documentclass[a4paper,11pt]{article}

\pdfoutput=1 

\usepackage{bbm}
\usepackage{jheppub} 
\usepackage[T1]{fontenc}
\usepackage{simplewick} 
\usepackage{makecell}	
\usepackage{tabu}		
\usepackage[makeroom]{cancel} 
\usepackage[normalem]{ulem} 
\usepackage[utf8]{inputenc}
\usepackage{amsmath,amsfonts,amsthm,amssymb} 

\newcommand{\p}{\partial}
\newcommand{\eq}{&\quad}

\newcommand{\lan}{\langle}
\newcommand{\ran}{\rangle}
\newcommand{\para}{\parallel}

\newcommand{\tr}{\text{tr}}
\newcommand{\diag}{\text{diag}}

\newcommand{\C}{\mathbb{C}}
\newcommand{\D}{\mathcal{D}}
\newcommand{\R}{\mathbb{R}}

\newcommand{\hx}{{\hat{x}}}

\newcommand{\al}{\alpha}
\newcommand{\be}{\beta}

\newcommand{\de}{\delta}

\newcommand{\ph}{\phi}
\newcommand{\g}{\gamma}

\newcommand{\la}{\lambda}

\newcommand{\rh}{\rho}

\newcommand{\thet}{\theta}

\newcommand{\x}{\xi}
\newcommand{\z}{\zeta}

\newcommand{\De}{\Delta}

\newcommand{\G}{\Gamma}

\preprint{UUITP-07/21}

\title{\boldmath Fusion of conformal defects in four dimensions}

\author{Alexander Söderberg}

\affiliation{Department of Physics and Astronomy, Uppsala University, Sweden}

\emailAdd{alexander.soderberg@physics.uu.se}

\makeatletter
\gdef\@fpheader{}
\makeatother

\abstract{We consider two conformal defects close to each other in a free theory, and study what happens as the distance between them goes to zero. This limit is the same as zooming out, and the two defects have fused to another defect. As we zoom in we find a non-conformal effective action for the fused defect. Among other things this means that we cannot in general decompose the two-point correlator of two defects in terms of other conformal defects. We prove the fusion using the path integral formalism by treating the defects as sources for a scalar in the bulk.}

\begin{document} 
	
\newtheorem{defin}{Definition}
\newtheorem{thm}{Theorem}
\newtheorem{cor}{Corollary}
\newtheorem{pf}{Proof}
\newtheorem{nt}{Note}
\newtheorem{ex}{Example}
\newtheorem{ans}{Ansatz}
\newtheorem{que}{Question}
\newtheorem{ax}{Axiom}

\maketitle

\section{Introduction}

Recently there have been a lot of research focused on higher dimensional (in dimensions greater than two) conformal field theories (CFT's) in the presence of a boundary (see e.g. \cite{BCFT:2005, BCFT:2009, BCFT:201200701, BCFT:201211}) or a defect (see e.g. \cite{DCFT:1601, DCFT:1706, DCFT:171208, DCFT:1807}). However, the literature regarding higher dimensional CFT's in the presence of several defects is scarce \cite{DD:16, DD:17, DD:1805, DD:1806}. We add to this literature this work where we consider two scalar Wilson lines in a free theory in four dimensions, and study the limit in which they intersect. This corresponds to a fusion of the two defects. \\

Fusion of defects have previously been studied in two dimensions \cite{2D:07120, 2D:15}, in supersymmetric theories \cite{SUSY:07, SUSY:11, SUSY:12}, and in topological field theories as well as in conformal nets using fusing categories \cite{MATH:02, MATH:1310, MATH:1312}. In this paper, we will provide three different examples of fusing two defects in four dimensional free theories using the path integral formalism. To our knowledge this has never been done before. The fusion in these three examples is done in the same way using the following method:

\begin{enumerate}
	\item Find the two-point function $\langle D_1D_2\rangle$ for the two defects. This should describe a non-perturbative Casimir effect between the defects, and will be given by the exponential of an integral over a two-point function of fields on each defect.
	\item Probe this correlator with a bulk field, $\langle D_1D_2\ph\rangle$, and study the fusing limit where the distance between the defects go to zero. This will be given by integrals over a two-point function of $\ph$ and a field from one of the defects. 
	\item Find an effective action for the fused defect. The difficulty here lies in identifying what kind of operators appear on the fused defect. In the examples we study, they are directional derivatives w.r.t. the distance vector between the two defects. The fused defect should satisfy
	\begin{equation}
	\begin{aligned}
	\langle D_1D_2\ph\rangle = \langle D_1D_2\rangle\langle D_f\ph\rangle \ .
	\end{aligned}
	\end{equation}
	\item Prove the fusing using the path integral formalism by treating the defects as sources for an operator in the bulk.\footnote{This might be more complicated for composite operators. For such operators we also have to be careful since they will generate a renormalization group (RG) flow on the defects.}
\end{enumerate}

Correlators with two conformal defects have previously been decomposed in conformal blocks corresponding to local operators in the bulk \cite{DD:16, DD:1806}. Only this block decomposition is known. It is therefore interesting to study whether two defects can be decomposed in another way, which would yield a bootstrap equation for the defect correlators. One might believe that another block decomposition would be in terms of other conformal defects. As we study explicit examples of fusion of conformal defects, we find that the fused defect is not conformal and thus in general there will not be a second decomposition in terms of conformal defects. \\

A higher dimensional CFT enjoy a $SO(d + 1, 1)$-symmetry in Euclidean space. We say that a flat (or spherical) $p$-dimensional defect is conformal if a field localized on it satisfy a $SO(p + 1, 1) \times SO(d - p)$-symmetry, where $SO(p + 1, 1)$ is the conformal symmetry along the defect, and $SO(d - p)$ is the group of rotations around the defect. The defect itself, as a $p$-dimensional operator, satisfy $SO(d - p)$-symmetry. \\

The first example we consider are two parallel Wilson lines $D_\pm$ separated by a distance $2R$ in a free theory in four dimensions. These are one-dimensional defects that does not carry any $SO(3)$-spin
\begin{equation}
\begin{aligned}
D_\pm = \exp \left( \la_\pm\int_{\R}dx\ph(x\hx_\para \pm R\hx_\perp^1) \right) \ , \quad \la_\pm \in \C \ .
\end{aligned}
\end{equation}

Here $\hx_\para$ is the unit vector parallel to the defect, and $\hx_\perp^1$ is one of the three unit vector orthogonal to the defect. In section \ref{Sec: Corr} we study the three-point correlator $\lan D_+D_-\ph\ran$ between the two defects probed with a bulk field $\ph$, and note that it is the same as a two-point correlator $\lan D_f\ph\ran$ between another defect and the same bulk field (times a non-perturbative Casimir effect from $\langle D_+D_-\rangle$ between the two defects). From this we deduce that the new defect $D_f$ is a fusion of the other two other defects. We find it to be given by
\begin{equation}
\begin{aligned}
D_f &= \exp\left(\sum_{n\geq 0}\frac{\la_-^n + (-1)^n\la_+^n}{n!}R^n\int_\mathbb{R}dx\p_{x_\perp^1}^n\ph(x)\right) \ .
\end{aligned}
\end{equation}

As the distance $R$ between the two defects $D_\pm$ goes to zero, the defect $D_f$ is conformal. However, as we zoom in and pick up perturbations in $R$, we find dimensionfull coupling constants which break the conformal symmetry. It is thus not possible to decompose the two-point correlator $\lan D_+D_-\ran$ in terms of conformal defects. \\

In section \ref{Sec: PI} we treat $D_\pm$ and $D_f$ as sources for $\ph$. We are then able to show that the path integral is the same regardless of whether we use $D_+$ and $D_-$ as a source, or $D_f$. This proves the fusion. \\

In section \ref{Sec: Ex: Circles on a line} and \ref{Sec: Ex: Concentric circles} we study two other examples of fusion in free theories. Here we consider two scalar Wilson loops (non-concentric and concentric) as opposed to Wilson lines. The defects are fused in a similar manner, and we can again prove it using the path integral formalism. Our main results in the paper are the fusions at equations \eqref{fusion}, \eqref{Fusion: Non-concentric} and \eqref{Fusion: Concentric}. \\

In appendix \ref{App: Blocks} we study the block decomposition of the two-point correlators of the circular defects considered in section \ref{Sec: Ex: Circles on a line} and \ref{Sec: Ex: Concentric circles}. This is done using the method of \cite{DD:16}. We find that all of the cross-ratios can be expressed in terms of each other, which means that this block decomposition does not need to be unique. \\

We conclude in section \ref{Sec: Conl} with some future aspects. Knowing the exponential form of two defects, we could possibly fuse them using the method presented in this paper. 

\section{Fusion of two Wilson lines, seen from defect correlators} \label{Sec: Corr}

We can write defects as exponentials, and one of the simplest examples is a scalar Wilson line in $d = 4$ dimensions \cite{DCFT:05, DCFT:1601}
\begin{equation}
\begin{aligned}
D = \exp\left(\la\int_{\R}dx\ph(x)\right) \ , \quad \De_\ph = 1 \ .
\end{aligned}
\end{equation}

Here $\la\in\C$ is a dimensionless constant. Using Wick's theorem we find 
\begin{equation} \label{Wick 1}
\begin{aligned} 
\langle D\rangle &= 1 + \sum_{n\geq 1}\frac{\la^n}{n!}\int_{\R}dx_1 ... \int_{\R}dx_n \langle\ph(x_1) ... \ph(x_n)\rangle \\
&= \sum_{n\geq 0}\frac{\la^{2n}}{2^nn!} \left( \int_{\R}d x_1\int_{\R}d x_2\langle\ph(x_1)\ph(x_2)\rangle \right)^n \\
&= \exp \left( \frac{\la^2}{2}\int_{\R}d x_1\int_{\R}d x_2\langle\ph(x_1)\ph(x_2)\rangle \right) \ .
\end{aligned}
\end{equation}

Correlators between local fields behave in the same way as in a homogenous CFT (without the defects). In the free theory it is given by
\begin{equation}
\begin{aligned}
\langle\ph(x_1)\ph(x_2)\rangle &= \frac{A_d}{|x_1 - x_2|^{2\De_\ph}} \ , \quad A_d = \frac{1}{(d - 2)S_d} = \frac{1}{4\pi^2} \ .
\end{aligned}
\end{equation}

Here $S_d$ is the area of a $(d - 1)$-dimensional sphere, and $\G_x \equiv \Gamma(x)$ is the Gamma-function. Integrals over fields on the same defects are divergent, so it is convenient to normalize the propagators by dividing with the one-point functions of the defects
\begin{equation}
\begin{aligned}
\langle D\rangle_N &\equiv \frac{\langle D\rangle}{\langle D\rangle} = 1 \ .
\end{aligned}
\end{equation}

\begin{figure} 
	\centering
	\includegraphics[width=0.5\textwidth]{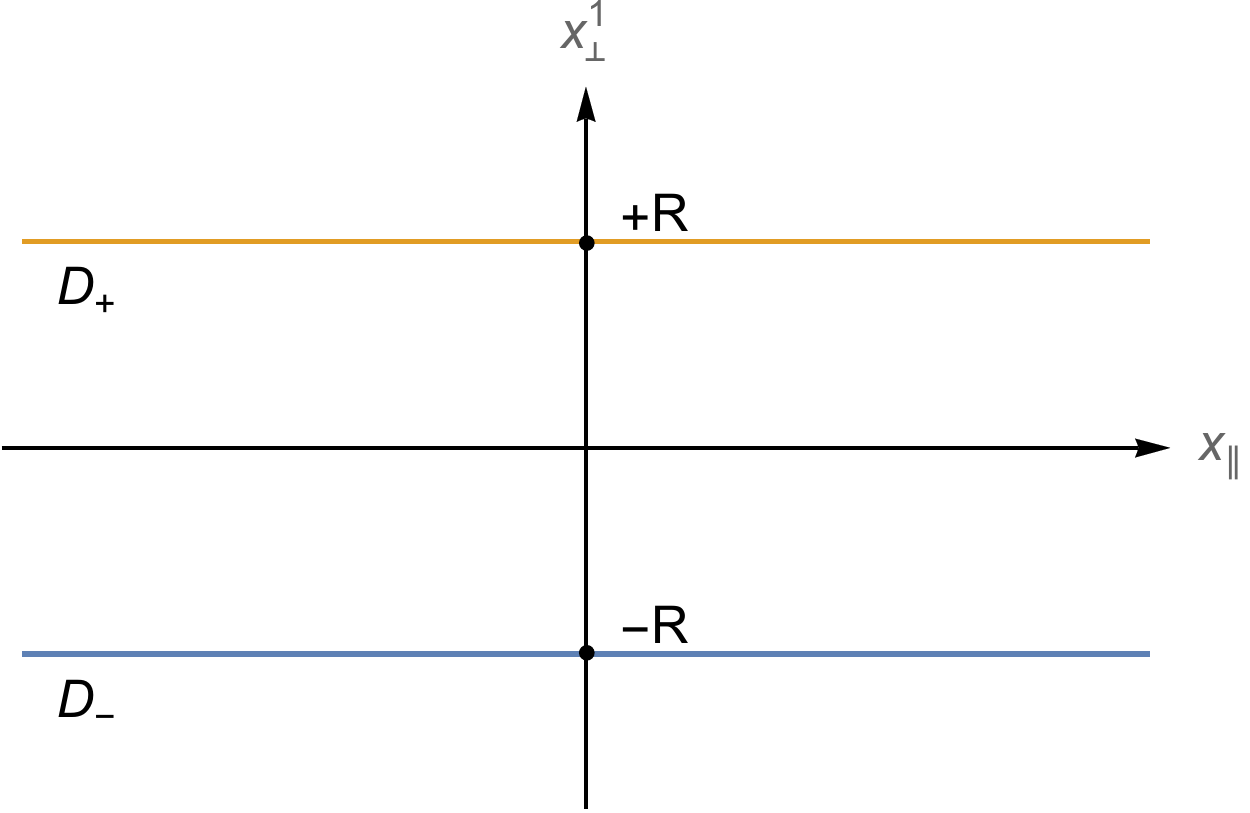}
	\caption{The two line defects are separated by a distance of $2R$.}
	\label{Fig: def}
\end{figure}

Now place two defects with a distance $2R > 0$ from eachother (see figure \ref{Fig: def})
\begin{equation} \label{Line def}
\begin{aligned}
D_\pm = \exp\left( \la_\pm\int_{\R}d x\ph(x\hx_\para \pm R \hx_\perp^1) \right) \equiv \exp\left( \la_\pm\int_{\R}d x\ph_\pm(x) \right) \ , \quad \la_\pm \in \C \ .
\end{aligned}
\end{equation}

Here $\hx_\para$ is the unit vector along the defect, and $\hx_\perp^1$ is one of the three unit vectors orthogonal to the defects. It will be convenient for us to use different notations depending on whether we integrate over fields on the same defect, or fields from different defects
\begin{equation} \label{Int 1}
\begin{aligned}
I_\pm \equiv \int_{\R}d x\int_{\R}d y\langle\ph_\pm(x)\ph_\pm(y)\rangle \ , \quad J \equiv \int_{\R}d x\int_{\R}d y\langle\ph_+(x)\ph_-(y)\rangle \ .
\end{aligned}
\end{equation}

\begin{figure} 
	\centering
	\includegraphics[width=0.5\textwidth]{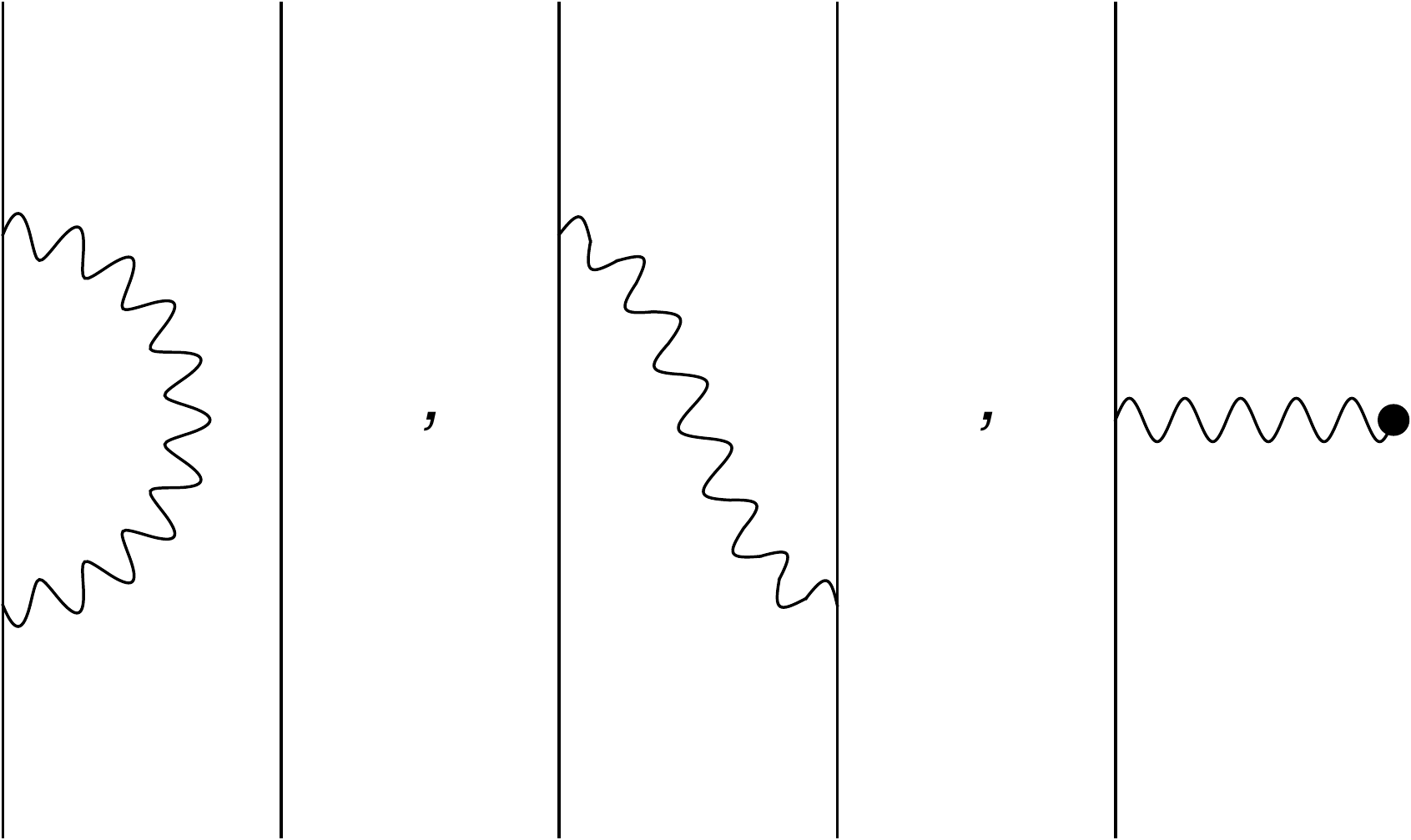}
	\caption{A diagrammatic depiction of the integrals $I_\pm \ , J$ and $K$.}
	\label{Fig: integrals}
\end{figure}

A diagrammatic representation of these integrals are in figure \ref{Fig: integrals}. The two-point function is given by
\begin{equation} \label{Wick 2}
\begin{aligned}
\langle D_+D_-\rangle &= \exp \left( \int_{\R} d x \left[ \la_+\ph_+(x) + \la_-\ph_-(x) \right] \right) \\
&= \sum_{n\geq 0}\frac{1}{n!}\langle \left( \int_{\R} d x \left[\la_+\ph_+(x) + \la_-\ph_-(x) \right] \right)^n \rangle \\
&= \sum_{n\geq 0}\frac{ \left( \la_+^2I_+ + \la_-^2I_- + 2\la_+\la_-J \right)^{n} }{2^nn!} = \exp \left( \frac{\la_+^2I_+ + \la_-^2I_-}{2} + \la_+\la_-J \right) \ .
\end{aligned}
\end{equation}

This yields the normalized correlator
\begin{equation}
\begin{aligned}
\langle D_+D_-\rangle_N &\equiv \frac{\langle D_+D_-\rangle}{\langle D_+\rangle\langle D_-\rangle} = e^{\la_+\la_-J} \ .
\end{aligned}
\end{equation}

The $J$-integral is found using a Julian-Schwinger parametrization and regularizing one of the defects such that it is of finite length $2L\gg 1$
\begin{equation} \label{Int J}
\begin{aligned}
J &= \lim\limits_{L\rightarrow\infty}\int_{-L}^{+L}d y\int_{\R}d x\int_{0}^{\infty}du\frac{e^{-u(x - y)^2 - 4uR^2}}{4\pi^2} = \lim\limits_{L\rightarrow\infty}\int_{-L}^{+L}d y\int_{0}^{\infty}du\frac{e^{-4uR^2}}{4\pi^{3/2}\sqrt{u}} \\
&= \lim\limits_{L\rightarrow\infty} \frac{L}{4\pi R} \ .
\end{aligned}
\end{equation}

This yields
\begin{equation}
\begin{aligned}
\langle D_+D_-\rangle_N &= \lim\limits_{L\rightarrow\infty}e^{\la_+\la_-L/(4\pi R)} \ .
\end{aligned}
\end{equation}

The non-perturbative dependence in $R$ describes a Casimir effect between the two defects. \\

Let us now probe the two-point correlator with a bulk scalar placed at
\begin{equation}
\begin{aligned}
z \equiv z_\para\hx_\para + z_\perp\hx_\perp^1 \ , \quad |z_\perp| > R \ .
\end{aligned}
\end{equation}

Note that in the fusion limit $R\rightarrow 0^+$ this scalar is not squeezed in between the two defects.
\begin{equation*} \hspace{-30px}
\begin{aligned}
\langle D_+D_-\ph(z)\rangle &= \sum_{n \geq 0}\frac{1}{n!}\langle\left(\int_{\R}d x\left[\la_+\ph_+(x) + \la_-\ph_-(x)\right]\right)^n\ph(z)\rangle \\
&= \sum_{n \geq 0}\frac{2n + 1}{(2n +1)!}\langle\left(\int_{\R}d x\left[\la_+\ph_+(x) + \la_-\ph_-(x)\right]\right)^{2n}\rangle\langle\int_{\R}d x\left[\la_+\ph_+(x) + \la_-\ph_-(x)\right]\ph(z)\rangle \\
&= \langle D_+D_-\rangle\left[\la_+K_+(z) + \la_-K_-(z)\right] \ ,
\end{aligned}
\end{equation*}
\begin{equation} \label{Int 2}
\begin{aligned}
K_\pm(z) &\equiv \int_{\R}d x\langle\ph_\pm(x)\ph(z)\rangle \ .
\end{aligned}
\end{equation}

A diagrammatic representation of this integral is in figure \ref{Fig: integrals}. This integral can again be solved with a Julian-Schwinger parametrization
\begin{equation}  \label{Int K}
\begin{aligned}
K_\pm(z) &= \int_{\R}d x\int_{0}^{\infty}du\frac{e^{-u(z_\para - x^2) - u(z_\perp \mp R)^2}}{4\pi^2} = \frac{1}{4\pi|z_\perp \mp R|} = \sum_{n\geq 0}\frac{(\pm R)^n}{4\pi z_\perp^{n + 1}}  \ .
\end{aligned}
\end{equation}

It yields the normalized correlator\footnote{Please note that we could have chosen a normalization where we divide with $\langle D_+D_-\rangle$ rather than $\langle D_+\rangle\langle D_-\rangle$. Such normalization would remove the non-perturbative Casimir effect.}
\begin{equation} \label{DDphi}
\begin{aligned}
\langle D_+D_-\ph(z)\rangle_N &\equiv \frac{\langle D_+D_-\ph(z)\rangle}{\langle D_+\rangle\langle D_-\rangle} = \lim\limits_{L\rightarrow\infty}e^{\la_+\la_-L/(4\pi R)}\sum_{n\geq 0}\frac{\la_+ + (-1)^n\la_-}{4\pi z_\perp^{n + 1}}R^n \ .
\end{aligned}
\end{equation}

Now we wish to study whether this correlator can be written in terms of a fused defect
\begin{equation}
\begin{aligned}
\langle D_+D_-\ph(z)\rangle_N &\overset{?}{=} \langle D_+D_-\rangle_N\langle D_f\ph(z)\rangle_N \ . 
\end{aligned}
\end{equation}

In order to understand this we need to find some kind of operators that the series in \eqref{DDphi} would correspond to. For this purpose, let us consider the following defect
\begin{equation}
\begin{aligned}
D_n &= \exp\left( \int_{\mathbb{R}}d x\p_\perp^n\ph(x\hx_\para) \right) \ , \quad \p_\perp \equiv \p_{x_\perp^1} = \p_R \ .
\end{aligned}
\end{equation}

If we probe its correlator with a bulk scalar\footnote{In principle we could instead consider $\lim\limits_{R\rightarrow 0^+}\p_\perp^nK_+$. However, as we will see in the next section, this yields the wrong fusion.}
\begin{equation} 
\begin{aligned}
\langle D_n\ph(z)\rangle &= \langle D_n\rangle\int_{\R}d x\langle\p_\perp^n\ph(x\hx_\para)\ph(z)\rangle = \langle D_n\rangle\lim\limits_{R\rightarrow 0^+}\p_\perp^nK_-(z) = \langle D_n\rangle\frac{(-1)^nn!}{4\pi z_\perp^{n + 1}} \ .
\end{aligned}
\end{equation}

The normalized correlator is thus
\begin{equation}
\begin{aligned}
\langle D_n\ph(z)\rangle_N &\equiv \frac{\langle D_n\ph(z)\rangle}{\langle D_n\rangle} = \frac{(-1)^nn!}{4\pi z_\perp^{n + 1}} \quad\Rightarrow\quad \frac{1}{4\pi z_\perp^{n + 1}} = \frac{(-1)^n}{n!}\langle D_n\ph(z)\rangle_N \ .
\end{aligned}
\end{equation}

Compare with (\ref{DDphi}) to find
\begin{equation}
\begin{aligned}
\langle D_+ D_-\ph(z)\rangle_N &= \lim\limits_{L\rightarrow\infty}e^{\la_+\la_-L/(4\pi R)}\sum_{n\geq 0}\frac{\la_- + (-1)^n\la_+}{n!}R^n\langle D_n\ph(z)\rangle_N \ .
\end{aligned}
\end{equation}

From this we can deduce that the defects have fused into a single defect
\begin{equation} \label{fusion}
\boxed{
\begin{aligned}
D_+D_- &= \lim\limits_{L\rightarrow\infty}e^{\la_+\la_-L/(4\pi R)}D_f \ , \\
D_f &= \exp\left(\sum_{n\geq 0}\frac{\la_- + (-1)^n\la_+}{n!}R^n\int_{\R}d x\p_\perp^n\ph(\vec{x})\right) \ .
\end{aligned}
}
\end{equation}

It is possible to check that this fusion holds for other probed bulk fields, e.g.
\begin{equation}
\begin{aligned}
\langle D_+D_-\ph^2(z)\rangle_N &= \langle D_+D_-\rangle_N\langle D_f\ph^2(z)\rangle_N \ .
\end{aligned}
\end{equation}

Now let us discuss the fused defect at \eqref{fusion}. In the strict fusing limit it is conformal, but as we zoom in and pick up perturbations in $R \ ,$ it has dimensionfull coupling constants, $R^n$, and is thus no longer conformal. Among other things this means that it is not possible for us to use this fusion for bootstrap purposes. 


\section{Proof of fusion through path integral formalism} \label{Sec: PI}

In order to prove that the fusion (\ref{fusion}) is correct we need to show that it does not matter for the path integral (which generates all of the correlators) whether we consider the two line defects, or the fused defect. The path integral is
\begin{equation}
\begin{aligned}
Z[J] &= \int\D\ph\exp\left[\int_{\R^d}d^dx\left(\frac{\left(\p\ph\right)^2}{2} + J\ph\right)\right] \ .
\end{aligned}
\end{equation}

Let us perform a partial integration on the source term and add a zero on the form \\ $\left(\p^{-1}J\right)^2 - \left(\p^{-1}J\right)^2$. Then we complete the square for $\ph$ and perform a partial integration on the residual $\left(\p^{-1}J\right)^2$-term
\begin{equation}
\begin{aligned}
Z[J] &= \int\D\ph\exp \left[ \int_{\R^d}d^dx \left( \frac{\left(\p\ph\right)^2 - 2\p^{-1}J\p\ph + \left(\p^{-1}J\right)^2 - \left(\p^{-1}J\right)^2}{2} \right) \right] \\
&= \int\D\ph\exp \left[ \int_{\R^d}d^dx \left( \frac{\left[\p\left(\ph - \p^{-2}J\right)\right]^2 + J\p^{-2}J}{2} \right) \right] \ .
\end{aligned}
\end{equation}

Here $\p^{-2}$ is the Green's function $G$. Perform the field redefinition
\begin{equation}
\begin{aligned}
\ph(x) \rightarrow \ph(x) + \left(\p^{-2}J\right)(x) \equiv \ph(x) + \int_{\R^d}d^dyG(x - y)J(y) \ .
\end{aligned}
\end{equation}

This gives us the normalized path integral
\begin{equation} \label{Path integral}
\begin{aligned}
\frac{Z[J]}{Z[0]} &= e^{\z[J]} \ , \quad \z[J] = \int_{\R^d}d^dx\int_{\R^d}d^dy\frac{J(x)G(x - y)J(y)}{2} \ .
\end{aligned}
\end{equation}

To prove that the fusion (\ref{fusion}) is correct we need to show that the above path integral for the two line defects is the same as that for the fused defect. We will do this by writing the defects as sources. For the two defects we write
\begin{equation} \label{Double defect source}
\begin{aligned}
J_b(x) &= \la_+\de(\vec{x}_\perp - \vec{R}) + \la_-\de(\vec{x}_\perp + \vec{R}) + \tilde{J}(x) \ , \quad \vec{R} = R\hx_\perp^1 \ .
\end{aligned}
\end{equation}

Here $\vec{x}_\perp$ is the vector orthogonal to the defects, and $\tilde{J}(x)$ is the actual source for $\ph$. For the fused defect we write
\begin{equation} \label{Fused defect source}
\begin{aligned}
J_f(x) &= \de(\vec{x}_\perp)\sum_{n\geq 0}\frac{\la_- + (-1)^n\la_+}{n!}R^n\p_{x_\perp^1}^n + \tilde{J}(x) \ .
\end{aligned}
\end{equation}

We want to show that
\begin{equation}
\begin{aligned}
\z[J_b] &= \z[J_f] \ .
\end{aligned}
\end{equation}

If we insert (\ref{Double defect source}) into (\ref{Path integral})
\begin{equation} \label{Path int, both def}
\begin{aligned}
\z[J_b] &= \int_{\R}d x_\para\int_{\R}d y_\para \left( \frac{\la_+^2 + \la_-^2}{2}G(\vec{s}_\para) + \la_+\la_-G(\vec{s}_\para + 2\vec{R}) \right) + \\
\eq + \int_{\R}d x_\para\int_{\R^d}d^dy\tilde{J}(y)\left[\la_+G(\vec{s}_\para + \vec{y}_\perp - \vec{R}) + \la_-G(\vec{s}_\para + \vec{y}_\perp + \vec{R})\right] + \\
\eq +  \int_{\R^d}d^dx\int_{\R^d}d^dy\frac{\tilde{J}(x)G(x - y)\tilde{J}(y)}{2} \ .
\end{aligned}
\end{equation}

Here $\vec{x}_\perp, \vec{y}_\perp$ are vectors orthogonal to the defects , and $\vec{s}_\para$ is the difference between the parallel coordinates along the defects 
\begin{equation}
\begin{aligned}
\vec{s}_\para \equiv (x_\para - y_\para)\hx_\para \ , \quad \vec{y}_\perp = y_i\hx_\perp^i \ , \quad \vec{x}_\perp = x_i\hx_\perp^i \ , \quad i\in\{1, 2, 3\} \ .
\end{aligned}
\end{equation}

We want to show that the functional (\ref{Path int, both def}) is the same as (\ref{Fused defect source}) inserted in (\ref{Path integral})
\begin{equation*} \hspace{-50px}
\begin{aligned}
\z[J_f] &= \int_{\R}d x_\para\int_{\R}d y_\para\sum_{m, n \geq 0}\frac{\la_+^2R^{m + n} + \la_-^2(-R)^{m + n} + 2\la_+\la_-(-R)^mR^n}{2m!n!} \left. \p_{x_\perp^1}^m\p_{y_\perp^1}^nG(\vec{s}_\para + \vec{x}_\perp - \vec{y}_\perp) \right|_{\vec{x}_\perp = \vec{y}_\perp = 0} + \\
\eq + \int_{\R}d x\int_{\R^d}d^dy\tilde{J}(y)\sum_{n\geq 0}\frac{\la_-R^n + \la_+(-R)^n}{n!}\p_{x_\perp^1}^n \left. G(\vec{s}_\para + \vec{x}_\perp - \vec{y}_\perp) \right|_{\vec{x}_\perp = 0}  + \\
\eq + \int_{\R^d}d^dx\int_{\R^d}d^dy\frac{\tilde{J}(x)G(x - y)\tilde{J}(y)}{2} \ .
\end{aligned}
\end{equation*}

This is the Taylor expansion of \eqref{Path int, both def} around $R = 0$. Here we used that the Green's function is symmetric w.r.t. $x$ and $y$
\begin{equation}
\begin{aligned}
G(x - y) = G(y - x) \quad\Rightarrow\quad G(\vec{s}_\para, \vec{x}_\perp - \vec{y}_\perp) = G(\vec{s}_\para, \vec{y}_\perp - \vec{x}_\perp) \ .
\end{aligned}
\end{equation}

This proves that the defect in \eqref{fusion} is indeed the fusion of the line defects in \eqref{Line def}.

\section{Fusion of two non-concentric Wilson loops} \label{Sec: Ex: Circles on a line}

\begin{figure} 
	\centering
	\includegraphics[width=0.5\textwidth]{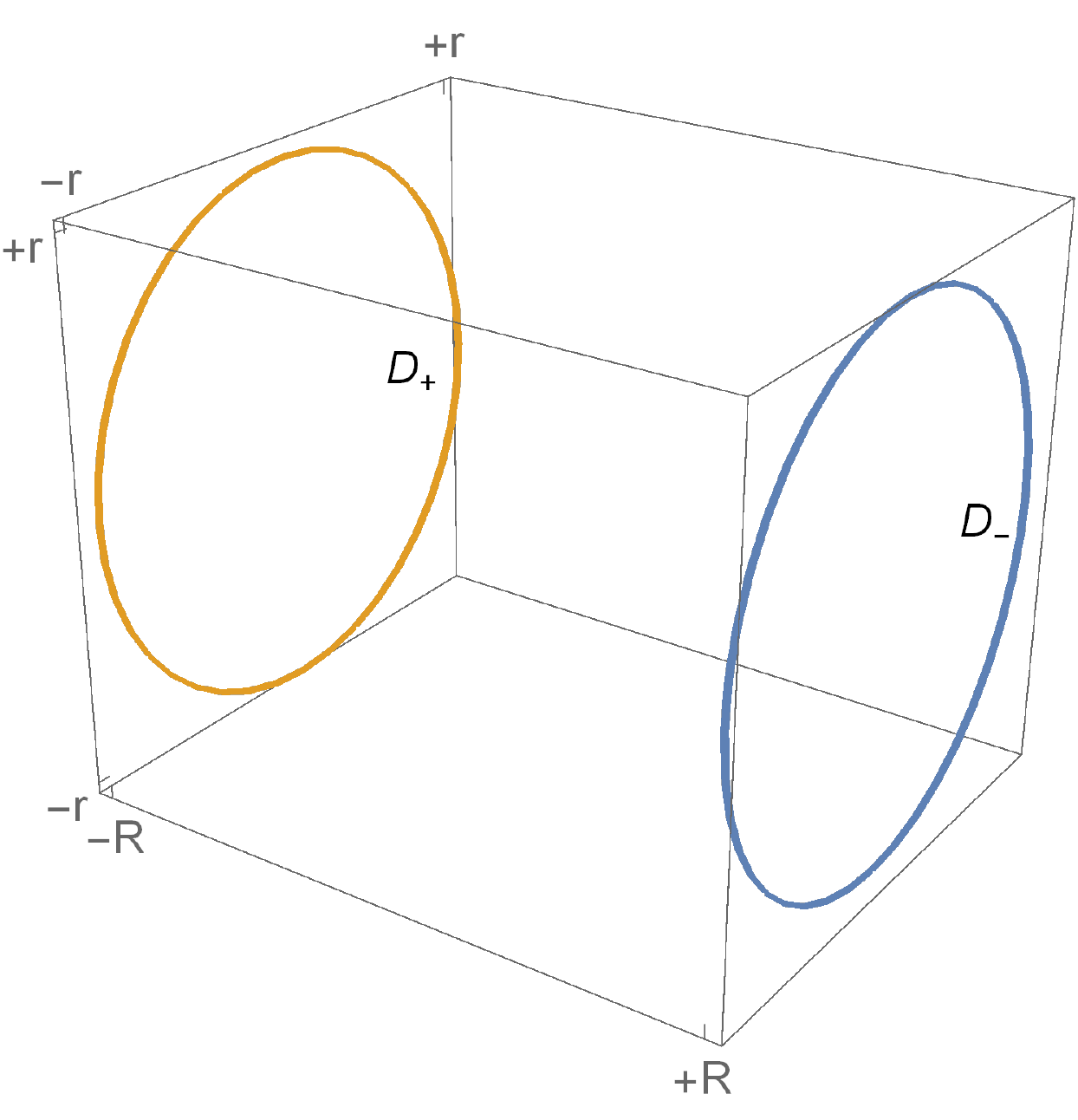}
	\caption{Two circular defects that are separated by a distance of $2R$.}
	\label{Fig: loop}
\end{figure}

In this section we will provide another example of fusion in a free theory. We will consider two scalar Wilson loops of radius $r$ in $d = 4$ dimensions at a distance $2R \equiv 2|\vec{R}| > 0$ from each other (see figure \ref{Fig: loop})
\begin{equation}
\begin{aligned}
D_\pm = \exp\left( \la_\pm\int_0^{2\pi}d\thet\ph(r(c_\thet\hat{x}_\para^1 + s_\thet\hat{x}_\para^2) \pm \vec{R}) \right) \equiv \exp\left( \la_\pm\int_0^{2\pi}d\thet\ph_\pm(\thet) \right) \ . 
\end{aligned}
\end{equation}

Here $c_\thet \equiv \cos(\thet) \ , s_\thet \equiv \sin(\thet)$ and $\hat{x}_\para^j \ , j\in\{1, 2\}$ are two of the four unit vectors. $\vec{R}$ is a vector orthogonal to both $\hat{x}_\para^j$. We can proceed in the same way as in section \ref{Sec: Corr}, encountering slightly different integrals corresponding to \eqref{Int 1} and \eqref{Int 2} (we probe the defect correlator with a bulk field at origo)
\begin{equation}
\begin{aligned}
J &= \int_{0}^{2\pi}d\thet_1\int_{0}^{2\pi}d\thet_2\langle\ph_+(\thet_1)\ph_-(\thet_2)\rangle \\
&= \frac{A_d}{2^{\De_\ph}}\int_{0}^{2\pi}d\thet_1\int_{0}^{2\pi}d\thet_2\int_0^\infty du\frac{u^{\De_\ph - 1}}{\G_{\De_\ph}}e^{-u(r^2(1 - c_{\thet_1 - \thet_2}) + 2R^2)} \\
&= \frac{\pi^2A_d}{2^{\De_\ph - 2}(r^2 + 2R^2)^{\De_\ph}}{}_2F_1 \left( \frac{\De_\ph}{2}, \frac{\De_\ph + 1}{2}; 1; \frac{r^4}{(r^2 + 2R^2)^2} \right) = \frac{1}{4R\sqrt{r^2 + R^2}} \ , \\
\end{aligned}
\end{equation}
\begin{equation}
\begin{aligned}
K_\pm &= \int_{0}^{2\pi}d\thet\langle\ph_\pm(\thet)\ph(0)\rangle = \int_{0}^{2\pi}d\thet\frac{A_d}{(r^2 + R^2)^{\De_\ph}} = \frac{1}{2\pi(r^2 + R^2)} \\
&= \sum_{n\geq 0}\frac{(-1)^nR^{2n}}{2\pi r^{2(n - 1)}} \ .
\end{aligned}
\end{equation}

We find the fusing to be
\begin{equation} \label{Fusion: Non-concentric}
\boxed{
\begin{aligned}
D_+D_- &= e^{\la_+\la_-/(4R\sqrt{r^2 + R^2})}D_f \ , \\ 
D_f &= \exp\left(\sum_{n\geq 0}\frac{\la_+ + \la_-}{(2n)!}R^{2n}\int_{0}^{2\pi}d\thet\nabla_R^{2n}\ph(\thet)\right) \ .
\end{aligned}
}
\end{equation}

Here $\nabla_R \equiv \vec{R} \cdot \vec{\nabla}$ is the directional derivative w.r.t. the vector $\vec{R}$. The fusion is proven in the same way as in section \ref{Sec: PI} using the path integral formalism and by treating the defects as sources for the fundamental scalar $\ph$ in the bulk.

\section{Fusion of two concentric Wilson loops} \label{Sec: Ex: Concentric circles}

\begin{figure} 
	\centering
	\includegraphics[width=0.5\textwidth]{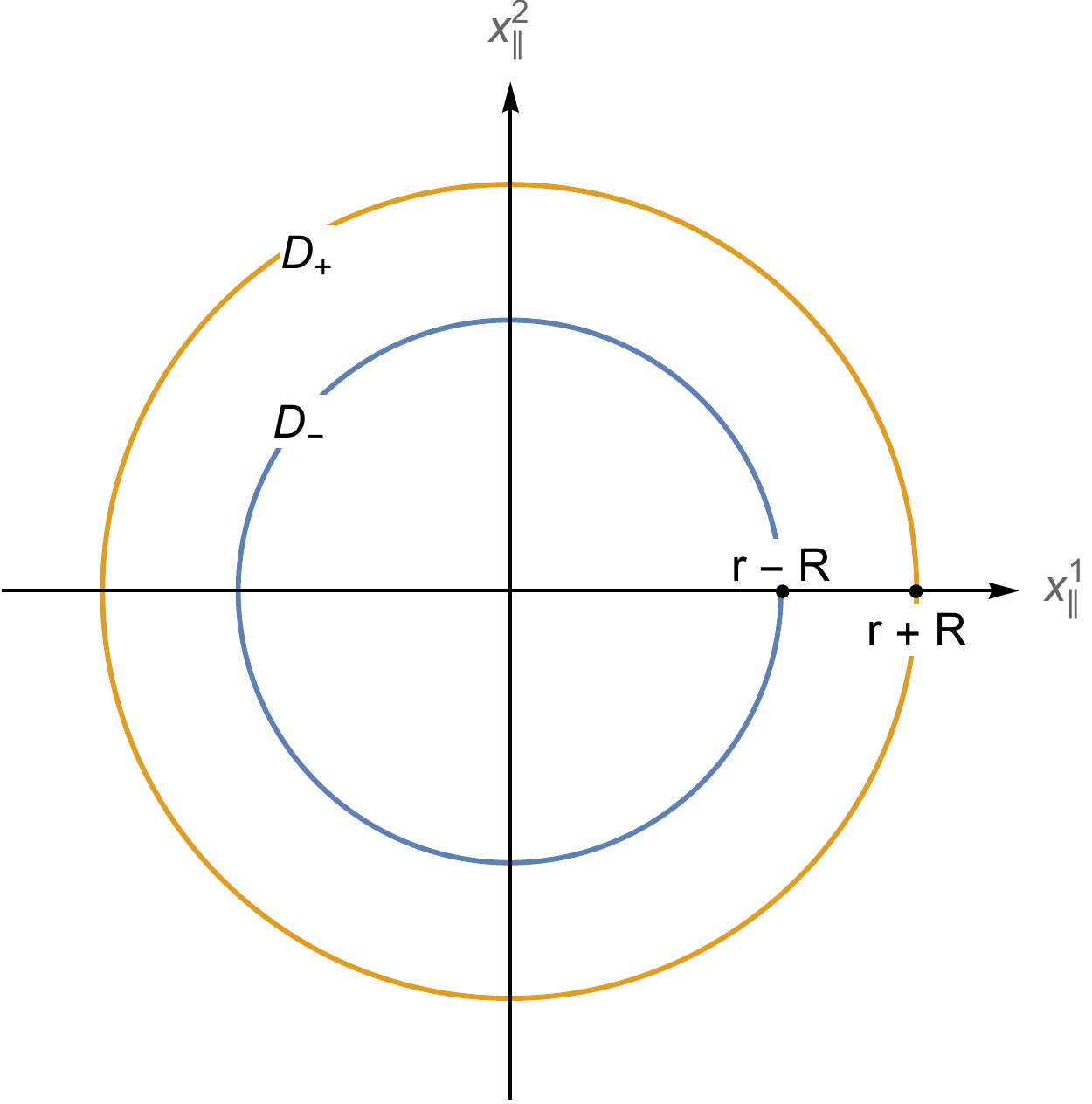}
	\caption{The concentric circular defects are separated by a distance of $2R$.}
	\label{Fig: concentric}
\end{figure}

The last example of fusion that we will study is between two concentric circles. We will again consider two scalar Wilson loops in $d = 4$ dimensions in a free theory, separated by a distance $2R$. Unlike the previous section the defects will be concentric (see figure \ref{Fig: concentric}), i.e. the Wilson loops will have different radii $r \pm R$
\begin{equation}
\begin{aligned}
D_\pm = \exp\left( \la_\pm\int_0^{2\pi}d\thet\ph \left( (r \pm R)(c_\thet\hat{x}_\para^1 + s_\thet\hat{x}_\para^2) \right) \right) \equiv \exp\left( \la_\pm\int_0^{2\pi}d\thet\ph_\pm(\thet) \right) \ . 
\end{aligned}
\end{equation}

The integrals we encounter in $\langle D_+D_-\rangle$ and $\langle D_+D_-\ph(0)\rangle$ (corresponding to \eqref{Int 1} and \eqref{Int 2}) are given by
\begin{equation}
\begin{aligned}
J &= \int_{0}^{2\pi}d\thet_1\int_{0}^{2\pi}d\thet_2\langle\ph_+(\thet_1)\ph_-(\thet_2)\rangle \\
&= \frac{A_d}{2^{\De_\ph}}\int_{0}^{2\pi}d\thet_1\int_{0}^{2\pi}d\thet_2\int_0^\infty du\frac{u^{\De_\ph - 1}}{\G_{\De_\ph}}e^{-u(r^2 + R^2 - (r^2 - R^2)c_{\thet_1 - \thet_2})} \\
&= \frac{\pi^2A_d}{2^{\De_\ph - 2}(r^2 + 2R^2)^{\De_\ph}}{}_2F_1 \left( \frac{\De_\ph}{2}, \frac{\De_\ph + 1}{2}; 1; \frac{(r^2 - R^2)^2}{(r^2 + R^2)^2} \right) = \frac{1}{4\pi R} \ , \\
\end{aligned}
\end{equation}
\begin{equation}
\begin{aligned}
K_\pm &= \int_{0}^{2\pi}d\thet\langle\ph_\pm(\thet)\ph(0)\rangle = \int_{0}^{2\pi}d\thet\frac{A_d}{(r \pm R)^{2\De_\ph}} = \frac{1}{2\pi(r \pm R)^2} \\
&= \sum_{n\geq 0}\frac{(\mp)^n(n + 1)R^n}{2\pi r^{n + 2}} \ .
\end{aligned}
\end{equation}

The fusion is very similar to the line defects in section \ref{Sec: Corr}
\begin{equation} \label{Fusion: Concentric}
\boxed{
\begin{aligned}
D_+D_- &= e^{\la_+\la_-/(4\pi R)}D_f \ , \\ 
D_f &= \exp\left(\sum_{n\geq 0}\frac{\la_- + (-1)^n\la_+}{n!}R^{n}\int_{0}^{2\pi}d\thet\p_R^n\ph(\thet)\right) \ .
\end{aligned}
}
\end{equation}

Here $\p_R^n$ are derivatives in the Radial direction. This fusion is proven in the same way as in section \ref{Sec: PI}.

\section{Conclusion} \label{Sec: Conl}

We have shown how several different defects in Gaussian models can be fused using one and the same method. This procedure should in principle work for other defects as well. Although we need to be able to express the defects as exponentials. It would be interesting to study whether this method works for fusion of defects with composite operators or higher dimensional defects. One has to be careful though, as such operator will generate an RG flow on the defect. \\

In this work we only considered free theories and it is thus worthwhile to study how the proposed method in this paper can be modified to work in an interacting theory (say with a quartic interaction in the bulk). This will affect the Wick contractions in equations \eqref{Wick 1}, \eqref{Wick 2} and the equation above \eqref{Int 2}. \\

Let us also comment a bit on the conformal block decomposition in \cite{DD:16}. The defect two-point correlators in section \ref{Sec: Ex: Circles on a line} and \ref{Sec: Ex: Concentric circles} can in principle be decomposed in these blocks, and there should be three number of independent cross-ratios. However, we find in appendix \ref{App: Blocks} that these cross-ratios are not independent from each other for two parallel circles of codiemension three in four spacetime dimensions.\footnote{As in figure \ref{Fig: loop}, but with different radii on the two circles.} This means that we cannot guarantee that this conformal block decomposition is unique.\footnote{I.e. we should treat appendix \ref{App: Blocks} as a step closer to an example of this block decomposition. Something that has not been done before to our knowledge.} \\

In the case of concentric Wilson loops studied in section \ref{Sec: Ex: Concentric circles}, we can also (in principle) decompose the defect two-point correlator using the method in \cite{DD:1806}. However, in order to use this method we first need to find the \textit{Harish-Chandra wave-functions} for codimension three defects ($N = 3$). \\

It would also be interesting to study whether fusion of defects can be applied to some more concrete physical examples. One such example may be the twist defect that appear in the context of Rényi entropy \cite{EE:04, EE:09, EE:1712}, where the so-called $c$-\textit{function} can be found through the fusion limit of two such defects \cite{EE:1506}. It is possible that this could be understood as a fused defect using the methods of this paper.

\section*{Acknowledgement}

The author is grateful to Marco Meineri and Emilio Trevisani for discussions regarding this project and for commenting on the manuscript. He is thankful to the organizers of the BOOTSTRAP 2019 workshop at Perimeter institute, where a majority of this project took place. AS is supported by Knut and Alice Wallenberg Foundation KAW 2016.0129.

\appendix

\section{Conformal block decomposition} \label{App: Blocks}

In this appendix we study the conformal block decomposition in \cite{DD:16} of the defect two-point correlators in section \ref{Sec: Ex: Circles on a line} and \ref{Sec: Ex: Concentric circles}. For this purpose we will assume the more general scalar Wilson loop
\begin{equation}
\begin{aligned}
D_\pm = \exp\left( \la_\pm\int_0^{2\pi}d\thet\ph(r_\pm(c_\thet\hat{x}_\para^1 + s_\thet\hat{x}_\para^2) \pm \rh\hx_\perp^1) \right) \ . 
\end{aligned}
\end{equation}

Here the two defects are of different radii $r_+$ and $r_-$, and $2\rh$ is the distance between the defects. We get the configuration in section \ref{Sec: Ex: Circles on a line} in the limit $r_+ \rightarrow r_- \equiv r$ with $\rh\hx_\perp^1 \equiv \vec{R}$, and that in section \ref{Sec: Ex: Concentric circles} in the limit $\rh\rightarrow 0$ with $r_\pm = r \pm R$. We need the following lightcone vectors to find the conformal cross-ratios
\begin{equation}
\begin{aligned}
P_\pm^1 &= (0, 0, 0, 0, 1, 0) \ , \\
P_\pm^2 &= (0, 0, 0, 0, 0, 1) \ , \\
P_\pm^3 &= \left( \frac{1}{r_\pm}, \frac{\rh^2}{r_\pm} - r_\pm, \pm\frac{\rh}{r_\pm}, 0, 0, 0 \right) \ .
\end{aligned}
\end{equation}

We are interested in the matrix
\begin{equation}
\begin{aligned}
M^{\al\be} = (\vec{P}_+^\al \cdot \vec{P}_-^\g)(\vec{P}_+^\g \cdot \vec{P}_-^\be)  = \diag\left( 1, 1, \x^2 \right)^{\al\be} \ , \quad \x = \frac{r_-^2 + r_+^2 - 4\rh^2}{2r_-r_+} \ .
\end{aligned}
\end{equation}

Here $\al, \be, \g \in\{1, 2, 3\}$, where summation over $\g$ is implicit. The three cross-ratios, $\eta_a$, with $a\in\{1, 2, 3\}$, are given by
\begin{equation}
\begin{aligned}
\eta_a = \tr(M^a) = 2 + \x^{2a} \ .
\end{aligned}
\end{equation}

As we can see, they can all be expressed in terms of another cross-ratio $\x$, and thus they are not independent from each other. This means that we cannot guarantee that the conformal block decomposition is unique. We can proceed to follow the procedure in \cite{DD:16} to find an ODE (w.r.t. $\x$) for the conformal blocks that we can possibly solve as a series expansion in $\x$. However, since we cannot determine whether this decomposition is unique, we will not write out any details on this. The interested reader may study the attached Mathematica file on arXiv.

\bibliographystyle{utphys}
\footnotesize
\bibliography{fusion}	
	
\end{document}